\documentclass[aps,prl,twocolumn,groupedaddress,showpacs]{revtex4}

\usepackage{graphicx}
\begin{document}
\title{Quantum thermal transport from classical molecular dynamics}

\author{Jian-Sheng Wang}
\altaffiliation{Also affiliated with 
Singapore-MIT Alliance, 4 Engineering Drive 3, Singapore 117576;
and Institute of High Performance Computing, 
1 Science Park Road, Singapore 117528.}
\affiliation{Center for Computational Science and Engineering, and
Department of Physics, National University of Singapore, Singapore
117542, Republic of Singapore} 

\date{22 June 2007}

\begin{abstract}
Using a generalized Langevin equation of motion, quantum ballistic thermal
transport is obtained from classical molecular dynamics.  This is possible
because the heat baths are represented by random noises obeying quantum
Bose-Einstein statistics.  The numerical method gives asymptotically exact
results in both the low-temperature ballistic transport regime and
high-temperature strongly nonlinear classical regime.  The method can be thought
of as a semi-classical approximation to the quantum transport problem.  A
one-dimensional quartic on-site model is used to demonstrate the crossover
from ballistic to diffusive thermal transport.
\end{abstract}


\maketitle


Many approaches have been used to study lattice heat transport in bulk
materials and nanostructures.  For bulk materials, the standard method is that
of Peierls based on Boltzmann equation for phonons \cite{peierls,rmp-review}.
For quasi-one-dimensional systems and nanojunctions, a variety of techniques
has been used, such as molecular dynamics (MD) \cite{lepri,mcgaughey}, mode-coupling
theory \cite{wang-mode}, nonequilibrium Green's function (NEGF) method 
\cite{ciraci,mingo,yamamoto,dhar,PRB-green,Mingo-PRB-negf}, Schr\"odinger
equation method \cite{michel}, quantum Langevin dynamics
\cite{haanggi,Dhar-heat-bath}, rigorous Boltzmann equations \cite{spohn}, etc.
One of the outstanding problems in heat transport is to reconcile the
ballistic nature at low temperatures and diffusive transport at high
temperatures.  As far as we know, the methods mentioned above work only in
either ballistic regime, or diffusive regime, but none correctly in both.

Molecular dynamics has the potential to be such a universal method for heat
transport.  However, since MD is based on classical Newtonian mechanics, the
quantum effect is completely absent.  Thus we can not expect that it is still
correct at low temperatures.  In fact, due to very high Debye temperatures for
carbon based materials, even $300\,$K is considered a low temperature.  The
kinetic theory of heat transport for phonons gives a formula for the thermal
conductivity as $\kappa = {1\over 3} c v l$, where $c$ is heat capacity, $v$
is sound velocity, and $l$ is mean free path.  The reduction of thermal
conductivity at low temperatures is mainly due to much reduced quantum heat
capacity $c$, but a classical MD can only produce a constant heat capacity.

Can we simulate a quantum system within MD?  At first sight, this seems
impossible, since classical dynamics can only produce classical results.  In
this paper, we show that the heat transport problem in junction systems can be
studied with a classical generalized Langevin dynamics using a quantum heat
bath derived from Bose-Einstein statistics.  Instead of the generic
Nos\'e-Hoover heat bath, it is essential to use the generalized Langevin dynamics
with memory kernel and colored noises to take care correctly the effect of the
baths.  The heat baths are modeled as infinite numbers of coupled harmonic
oscillators.  A remarkable feature of the proposed dynamics is that it
reproduces the quantum ballistic results at low temperature when nonlinearity
can be neglected, as well as gives a correct high-temperature, strongly
nonlinear result.  This appears to be the only method that is numerically
exact in both limits.  Although the classical and quantum generalized Langevin
equations are well-known, it is somewhat surprising that they are seldom used
in molecular dynamics.  In fact, they have much better properties with respect
to heat baths; we advocate their use for thermal transport problems.  Our
method is inspired by the NEGF approach
\cite{PRB-green} to heat transport and also the quantum Langevin approach
\cite{Dhar-heat-bath} to the same problem.

In the rest of the paper, we first introduce the model and give the equations
involved.  We then compare the MD results with Landauer formula and with
the nonlinear NEGF results.  We treat a one-dimensional (1D)
quartic nonlinear onsite model, in which we have seen ballistic transport at
temperatures below $200\,$K, and diffusive transport about $1000\,$K for
lattice sizes up to 4096.

The general setup of our system consists of a central junction region
connected to two semi-infinite harmonic lattices which serve as leads.  The
Hamiltonian of the system is
\begin{equation}
{\cal H} = \!\!\!\!\!\sum_{\alpha=L,C,R}\!\!\!\!\!H_\alpha  + (u^L)^T V^{LC} u^C + (u^C)^TV^{CR} u^R + H_n,
\end{equation}
where $H_{\alpha} = \frac{1}{2} {(\dot{u}^\alpha)}^T \dot{u}^\alpha +
\frac{1}{2} {(u^\alpha)}^T K^\alpha u^\alpha$, $u^\alpha$ is a column vector
consisting of all the displacement variables in region $\alpha$ ($=$ L, C, R),
and $\dot{u}^\alpha$ is the corresponding conjugate momentum.  The superscript
$T$ stands for matrix transpose.  We have chosen a renormalized displacement
$u_j = \sqrt{m_j}\, x_j$ where $m_j$ is the mass associated with $j$-th degree
of freedom, $x_j$ is the actual displacement having the dimension of length.
$K^\alpha$ is the spring constant matrix and $V^{LC}=(V^{CL})^T$ is the
coupling matrix of the left lead to the central region; similarly for
$V^{CR}$.  The equations of motions are of the form
\begin{eqnarray}
\ddot{u}^C &=& - K^C u^C +F_n(u^C) - V^{CL}u^L - V^{CR} u^R,\\
\ddot{u}^L &=& - K^L u^L - V^{LC}u^C,\\
\ddot{u}^R &=& - K^R u^R - V^{RC}u^C.
\end{eqnarray}
The heat-bath degrees of freedom $u^L$ and $u^R$ can be eliminated by solving
them in terms of the central variables and initial conditions, given, e.g.,
for the left lead:
\begin{eqnarray}
u^L(t) &=& \int_{t_0}^t g(t,t') V^{LC} u^C(t')\, dt' \nonumber \\
   & & +\, {\partial g(t,t_0) \over \partial t_0} u^L(t_0)  - g(t,t_0) \dot{u}^L(t_0),
\end{eqnarray} 
where $g(t,t')$ is the time-domain retarded surface Green's function of the
left lead obtained by the solution of
\begin{equation}
{\partial^2 g(t,t') \over \partial t'^2} + g(t,t') K^L = -\delta(t-t') I,
\end{equation}
with the condition $g(t,t') = 0$ if $t-t' \leq 0$.

Substituting the formal solutions of the leads into the central region, we
obtain the following generalized Langevin equation
\cite{Dhar-heat-bath,G-Langevin,peter-review} for the central part of the
degrees of freedom:
\begin{equation}
\ddot{u}^C = -K^C u^C + F_n(u^C\!)\! - \!\!\int_{t_0}^t\!\! \Sigma(t, t')u^C(t') dt' 
+ \xi_L + \xi_R,
\label{langevin}
\end{equation}
where $F_n$ is the nonlinear force, $\Sigma$ is the retarded self-energy of
the leads, $\Sigma = \Sigma_L +\Sigma_R$, as used in the NEGF
calculation, but in the time domain; $\Sigma_L = V^{CL} g V^{LC}$.  A
similar equation holds for the right lead $\Sigma_R$ using the right lead
surface Green's functions.  Contribution from the left lead due to the initial
conditions is
\begin{equation}
\xi_L(t) = V^{CL} \left( g(t,t_0) \dot{u}^L(t_0) 
- \frac{\partial g(t, t_0)}{\partial t_0} u^L(t_0) \right).
\label{random-noise}
\end{equation}
The expression for the right lead $\xi_R$ is analogous.  The initial time
$t_0$ will be set to $-\infty$.  Using the concept of adiabatic switch-on, at
time $-\infty$, the three subsystems, left lead, central region, and right
lead, are decoupled and the leads are in respective thermal equilibrium.  We
turn Eq.~(\ref{langevin}) into a stochastic differential equation by requiring
that $u^L(t_0)$ and $\dot{u}^L(t_0)$ are random variables.

So far we have treated the system as a classical system.  However, at this
point, we'll make a departure and treat the leads quantum-mechanically.  Since
the lead system is linear, the classical equation of motion and quantum
Heisenberg equation of motion are identical.  At time $t_0 \to -\infty$, the
leads are isolated.  We assume that the leads obey a quantum Bose-Einstein
statistics.  This induces a random variable $\xi_L(t)$ having zero mean,
$\langle \xi_L(t) \rangle = 0$, and the following correlation matrix
\begin{eqnarray}
\langle \xi_L(t) \xi_L(t')^T \rangle  = 
V^{CL} \Bigl(\dot{g}(t,t_0)\langle u^L(t_0) u^L(t_0)^T \rangle \dot{g}(t',t_0)^T \nonumber\\
 -\,\dot{g}(t,t_0)\langle u^L(t_0) \dot{u}^L(t_0)^T \rangle g(t',t_0)^T \qquad  \nonumber\\
   -\,g(t,t_0)\langle \dot{u}^L(t_0) u^L(t_0)^T \rangle \dot{g}(t',t_0)^T\qquad \nonumber\\
 \!\!\! + \, g(t,t_0) \langle \dot{u}^L(t_0) \dot{u}(t_0)^T\rangle g(t',t_0)^T \Bigr) V^{LC}. 
\end{eqnarray}
For a sensible heat bath, the correlation should be time translationally
invariant and independent of $t_0$.  Indeed, great simplification can be done
if we use the eigenmode representation for the matrix $g$:
\begin{equation}
g(t,t') = S^T g^d S, \quad g_j^d = - \theta(t\!-\!t') 
{\sin \omega_j (t\!-\!t') \over \omega_j },
\end{equation}
where $S$ is the orthogonal matrix that diagonalizes $K^L$, $S K^L S^T =
\Omega^2$, $\Omega^2$ is a diagonal matrix with diagonal elements
$\omega_j^2$; $\omega_j$s are the positive eigen frequencies.  Substituting
this result into the correlation expression, also using the quantum
equilibrium correlation values for $\langle u u^T \rangle$, $\langle \dot{u}
u^T \rangle$, $\langle u \dot{u}^T \rangle$, and $\langle \dot{u} \dot{u}^T
\rangle$, we obtain
\begin{equation}
\langle \xi_L(t) \xi_L(t')^T \rangle = V^{CL} S^T D S V^{LC},
\end{equation}
where $D$ is a diagonal matrix with elements $D_j = \bigl(2f(\omega_j) +
1\bigr) \frac{\hbar}{2\omega_j} \cos\omega_j(t-t') +
\frac{\hbar}{2i\,\omega_j} \sin\omega_j(t-t')$.  $f(\omega) =
\bigl[\exp(\beta_L \hbar \omega) -1\bigr]^{-1}$ is the Bose-Einstein
distribution function at the temperature of the left lead.

The appearance of the imaginary number in the last term in $D$ seems ominous,
as we cannot simulate a heat bath with imaginary correlation. The imaginary
part comes from the fact that in quantum mechanics, $\xi(t)$ and $\xi(t')$ are
non-commuting, and the product of the two is not a Hermitian operator.  Such a
difficulty can be easily overcome if we use a symmetrized correlation
$\frac{1}{2} \bigl\langle \xi_j(t) \xi_l(t') + \xi_l(t') \xi_j(t)
\bigr\rangle$.  This amounts to interchanging $t$ and $t'$ and taking the
transpose.  The final effect is simply to drop the imaginary term.

Then the question arises that such a treatment will not give correctly
the quantum results.  It turns out that it causes no problem, at least for the
expression of heat current.  We can show rigorously that, with the
symmetrized heat baths, we reproduce exactly the Landauer result with
Caroli-formula as the transmission coefficient.  However, the symmetrization
does have a consequence to the quantum heat-current fluctuations.

Using the (surface) density of states, the expression can be further
simplified to get a rather compact result for the spectrum of the noises
\cite{Dhar-heat-bath},
\begin{equation}
\tilde{F}[\omega]=\!\int_{-\infty}^\infty\!\!\!\!
\bigl\langle \xi_L(t) \xi_L^T(0)\bigr\rangle  e^{i\omega t} dt 
= \Bigl(f_L(\omega) + \frac{1}{2}\Bigr)
\hbar \Gamma_L[\omega],
\end{equation}
where $\Gamma_L[\omega] = i \bigl( \Sigma_L[\omega] - \Sigma_L[\omega]^\dagger
\bigr) = -2\, {\rm Im}\, V^{CL} g[\omega] V^{LC}$.  The spectrum function
$\tilde{F}[\omega]$ is even in $\omega$ and is a symmetric matrix.  Classical
limit is obtained if we take $\bigl(f(\omega) + 1/2\bigr)\hbar \approx k_B
T_L/\omega$, where $k_B$ is the Boltzmann constant and $T_L$ is 
the temperature of left lead.
 
The thermal current in steady state can be computed in several equivalent ways:
\begin{eqnarray}
I_L = -I_R = - \bigl\langle {d H_L \over dt} \bigr\rangle = 
\langle (\dot{u}^L)^T V^{LC} u^C \rangle  \qquad\qquad\nonumber \\
\label{eqIdef}
=-\langle u^C(t)^T \dot{B}(t) \rangle =
 \langle \dot{u}^C(t)^T B(t) \rangle, \quad
\end{eqnarray}
where $B(t) = -\int_{t_0}^t \Sigma_L(t,t') u^C(t') dt' + \xi_L(t)$.

The stochastic differential equation, Eq.~(\ref{langevin}), can be solved
numerically in a straightforward way.  Both the memory function (retarded
self-energy $\Sigma$) and noise spectrum $\tilde{F}$ can be obtained through
the surface Green's function $g$.  Efficient recursive algorithms exist for
the solution of $g$ \cite{PRB-green,surface-green}.  A set of past
coordinates, $u^C(t)$, needs to be stored, in order to perform a numerical
integration due to the self-energy.  We can use a simple rectangular rule for
the integration.  The random noises can be generated using a spectrum method
\cite{color-noise-book}.  Let the discrete Fourier transform of $\xi(t)$ be
$\eta_k = \eta^{*}_{-k} = a_k + i b_k$, $k = -M/2, \cdots, -1, 0, 1, \dots,
M/2-1$; where $M$ is the number of sampling points in the discrete Fourier
transform. Then the noises can be generated by taking real numbers $a_k$ and
$b_k$ ($k>0$) as independent Gaussian random numbers with zero mean and
variance $\frac{1}{2} \tilde F[\omega_k] h M$, where $h$ is the integration
step size, $\omega_k = 2\pi k/(h M) $.  The noise values at required times are
obtained by an inverse fast Fourier transform as $\xi(t=h l) = \frac{1}{hM}
\sum_{k} \eta_k \exp(-i2\pi l k /M)$.  The numerical integration of
Eq.~(\ref{langevin}) is not substantially more expensive than standard
MD. This is because the forces are usually short-ranged; we only need to do
the extra work for these sites that are directly connected to the leads.  Note
that the matrix elements of $V^{CL}$ and $V^{CR}$ are zero except those that
have a direct connection between the center and leads.  The computational
complexity becomes more favorable as the system becomes larger.

To illustrate the general method, we consider a simple 1D model with a quartic
on-site potential ($\phi^4$ model).  Such a model is known to have diffusive
transport in the classical limit \cite{bambiHu}.  The equation of motion is
given by
\begin{equation}
 \ddot{u}_j =  K u_{j-1} - (2K + K_0) u_j + K u_{j+1} - \mu_j u_j^3,
\end{equation}
where the nonlinear term is nonzero only in the central region, i.e., $\mu_j =
\mu$ if $1 \leq j \leq N$ and $\mu_j = 0$ otherwise.  The required surface
Green's function can be obtained analytically in frequency domain as
$g[\omega] = -\lambda/K$, where $\lambda$ is the root of the quadratic
equation, $K \lambda^{-1} + (\omega + i 0^{+} )^2 - 2K - K_0 + K \lambda = 0$,
such that $|\lambda| < 1$.  We use the following expression for the heat
current \cite{lepri},
\begin{equation}
\label{eqImd}
I^{{\rm MD}} = \frac{K}{2} \bigl\langle (\dot{u}_j + \dot{u}_{j+1}) 
(u_j - u_{j+1}) \bigr\rangle.
\end{equation}
Due to energy conservation along the chain, one can show that
Eq.~(\ref{eqImd}) is equal to that defined by Eq.~(\ref{eqIdef}) in steady
state.

\begin{figure}
\includegraphics[width=\columnwidth]{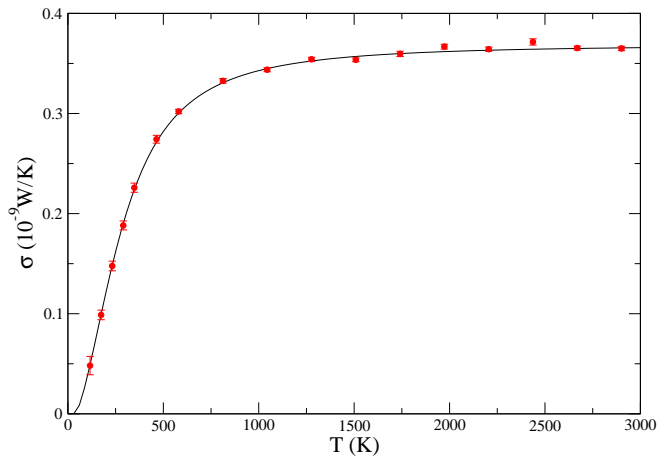}
\caption{\label{fig-1}Thermal conductance $\sigma$ for the 1D onsite model
without the nonlinear interaction, with spring constant 
$K = 1.0\,$eV/(amu\AA$^2$), $K_0 = 0.1 K$.  The smooth curve is the Landauer formula result,
while the symbols are MD results with a size $N=8$.  The time-step $h =
10^{-16}$s and $5\times 10^8$ MD steps each are used.}
\end{figure}

We now present our numerical results.  First, when there is no nonlinear
interaction, $\mu_j \equiv 0$, the heat current can be computed exactly
through the Landauer/Caroli formula, $I_L = \frac{1}{2\pi} \int_{0}^{\infty}
d\omega\, \hbar \omega\, {\rm Tr}(G^r \Gamma_L G^a \Gamma_R) (f_L - f_R)$,
where $G^r = (G^a)^\dagger = \bigl( (\omega + i0^+)^2 -K^C -\Sigma
\bigr)^{-1}$.  The molecular dynamics with the quantum heat bath reproduces
this result exactly.  In Fig.~\ref{fig-1}, we present the comparison of MD and
the exact curve.  The conductance is defined by
\begin{equation}
\sigma = \lim_{ T_L \to T_R} { I_L \over T_L - T_R}.
\end{equation}
A numerical finite-difference with 10-percent above or below the average
temperature is used.  Within MD statistical errors (computed from statistical
fluctuations of multiple runs), the agreement is perfect.  For the ballistic
transport, the thermal conductance is independent of the lengths $N$ of the
system.

\begin{figure}
\includegraphics[width=\columnwidth]{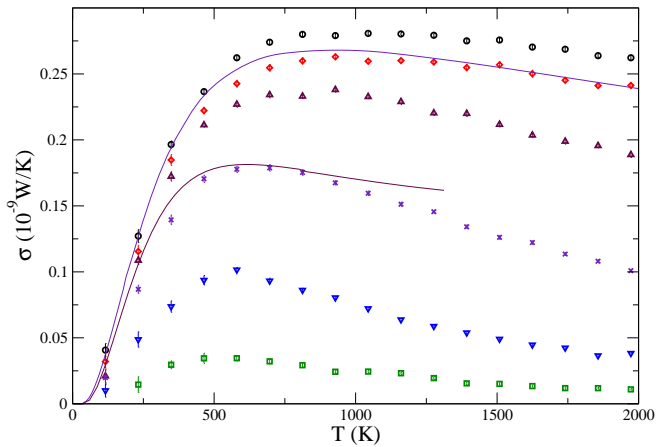}
\caption{\label{fig-2}Thermal conductance $\sigma$ for the 1D onsite model
with the nonlinear interaction $\mu = 1\,$eV/(amu$^2$\AA$^4$), spring constant
$K = 1.0\,$eV/(amu\AA$^2$), $K_0 = 0.1 K$.  The smooth curves are the NEGF
results for sizes $N=4$ and 32, respectively, while the symbols are MD results
with size $N$ from 4, 16, 64, 256, 1024, to 4096, from top to bottom.  The
time-step $h = 10^{-16}$s and $10^8$ MD steps each are used.}
\end{figure}

A nontrivial result is obtained when the system has nonlinear interactions.  This is
presented in Fig.~\ref{fig-2}.  With a nonlinear strength of $\mu=1$
[eV/(amu$^2$\AA$^4$)], we obtain quantitatively correct picture of ballistic
transport at low temperatures and small sizes ($\sigma \propto N^0$) 
and diffusive transport
at high temperatures and large sizes ($\sigma \propto 1/N$).  The
low-temperature results can be compared with the NEGF
ones.  This is presented as smooth curves in Fig.~\ref{fig-2}. The
NEGF results are obtained with a mean-field
approximation to the self-energies \cite{PRB-green}. The Green's functions
are iterated in equilibrium and the conductance is calculated with an
approximate formula for the transmission, $\tilde{T}[\omega] = \frac{1}{2}
{\rm Tr} \bigl\{ G^r (\Gamma_L + \frac{1}{2}\Gamma_n ) G^a \Gamma_R \bigr\} +
\frac{1}{2} {\rm Tr} \bigl\{ G^a \Gamma_L G^r (\Gamma_R + \frac{1}{2}\Gamma_n
) \bigr\}$, where the nonlinear effect is reflected in the extra nonlinear
self-energy, $\Gamma_n = i (\Sigma_n^r - \Sigma_n^a)$.  The MD and NEGF
results agree with each other at the low-temperature side very well.  Clearly,
the nonlinear NEGF results are not exact at high temperatures.  Thus the
deviation between MD and NEGF is understandable.  
If classical heat baths are used, then as the temperature decreases,
the thermal conductance increases monotonically to a size-independent ballistic
value of $(\omega_{\rm max}\! -\! \omega_{\rm min})k_B/(2\pi)$, where
$\omega_{\rm max} - \omega_{\rm min}$ is the phonon band width. 
At the intermediate range of temperatures, no
reliable methods exist that can be compared with the quantum MD results.
Thus, in this difficult temperature range, the MD results are the only numbers
to offer.  
Whether we see ballistic or diffusive
transport in a given temperature is determined by the mean free path of the
phonons in comparison with the system size $N$.  From the data in
Fig.~\ref{fig-2}, we can judge that the mean free path is about $10^3$ lattice
spacings in temperature range of $1000\,$K.

The dynamics also gives correctly the quantum average energy and quantum heat
capacity (say, with equal temperatures for the two leads).  
This is consistent with the fact that quantum conductance is
calculated correctly.  In classical simulation, the average kinetic energy
gives the local temperature of the system, $\langle \dot{u}^2_j\rangle = k_B
T$.  However, this is not true in our dynamics and the kinetic energy is
several times larger than implied by the equipartition theorem.  Interestingly,
in the limit of high temperatures of several thousand Kelvin, the
equipartition theorem is restored.

In summary, we showed that a generalized Langevin dynamics as a classical
stochastic differential equation can reproduce quantum ballistic transport if
the heat baths follow the quantum prescription.  This is achievable because
there is very little difference between a quantum and classical system if the
system is linear.  The dynamics is such that it smoothly crosses over to the
classical regime.  Thus the method produces correctly results both in the
quantum ballistic limit and classical diffusive limit.  We have applied the
method to a simple 1D onsite model.  Clearly, it is of general applicability.
For example, we can use the approach to study ballistic and diffusive thermal
transport in carbon nanotubes and graphene ribbons.  We can also study the
nonlinear effect in interfaces.  The present method opens new way for studying
quantum transport and nonlinearity.

The author thanks Jingtao L\"u, Jian Wang and Imam Makhfudz for discussions.
The computations were performed on the clusters of the Center for
Computational Science and Engineering and of Singapore-MIT Alliance, 
as well as on IBM cluster of Institute of High Performance Computing.
This work is supported in part by a Faculty Research Grant of the
National University of Singapore.

\end{document}